\documentclass[amsmath,amssymb,showpacs,floatfix,nofootinbib,prb]{revtex4}

\usepackage{graphicx}
\usepackage{dcolumn}
\usepackage{bm}
\usepackage{color}
\usepackage{extarrows}
\usepackage{enumitem}
\usepackage{threeparttable}
\usepackage{relsize} 
\newcommand{\Gl}{Eq.}
\newcommand{\gl}{Eq.}

\newcommand{\gls}{Eqs.}
\newcommand{\secname}{Section~} 

\DeclareGraphicsExtensions{.eps,.jpg,.tif,.pdf,.png}
\usepackage{hyperref}
\newcommand{\ib}[1]{{\color{black}#1}}
\newcommand{\figname}{Figure~} 
\newcommand{\figsname}{Figures~}
\begin{document}

\title[Why Asymmetric Molecular Coupling to Electrodes Cannot Be at Work in Real Molecular Rectifiers]{Why Asymmetric Molecular Coupling to Electrodes Cannot Be at Work in Real Molecular Rectifiers}
\author{Ioan B\^aldea}
\affiliation{%
Theoretical Chemistry, Heidelberg University, Im Neuenheimer Feld 229, 
D-69120 Heidelberg, Germany}

\begin{abstract}
Every now and then one can hear in the molecular electronics community that 
asymmetric couplings ($\Gamma_{s} \neq \Gamma_{t}$) of the dominant level (molecular orbital) to electrodes ($s$ and $t$)
which typically have shapes different of each other 
may be responsible for current rectification observed in experiments. Using a general single level model going beyond 
the Lorentzian transmission limit, in this work we present a rigorous demonstration that this is not the case.
In particular, we deduce an analytical for the bias ($V$) driven shift of the level energy $\delta \varepsilon_{0}(V)$
showing that $\delta \varepsilon_{0}(V)/V$ scales as $\Gamma_t/W_t - \Gamma_s/W_s$, which is merely a tiny quantity
because the electrode bandwidths $W_{s,t}$ are much larger than $\Gamma_{s,t}$. 
This result invalidates a previous, never-deduced
formula in use in some previous publications that neither could be justified theoretically nor is supported by experiment.
To the latter aim, we present new experimental evidence adding to that already inferred in earlier analysis.
\end{abstract}
\maketitle
\section{Introduction}
\label{sec:intro}
Current rectification ($RR \equiv I(V)/\vert I(-V)\vert \neq 1$)
using single molecule devices, a topic pioneered
by Aviram and Ratner,\cite{Aviram:74} continues to represent a major topic of molecular
electronics.\cite{Yaliraki:98a,Metzger:99,Nitzan:01,Tao:09,Whitesides:10b,Painelli:11,Jia:13,Wickramasinghe:13,Whitesides:14b,Metzger:15,Ratner:15b,Guo:16b,Trasobares:16,Nijhuis:17a,Thong:18,Metzger:18,Whitesides:19}
The present work is motivated by a 
confusion that persists on the physical origin of this phenomenon.
It is generated by the fact that
electrodes used to fabricate molecular junctions (planar substrate $s$ and more or less sharp tip $t$) have often shapes different of each other.
So, merely guided by naive intuition, one was often tempted to claim that the asymmetry in the measured current-voltage curves
is just a manifestation of electrodes' asymmetry. 
This issue has been addressed in a series of
publications in the past.\cite{Metzger:01b,Baldea:2012b,Baldea:2015a,Baldea:2015c,Baldea:2015g,Ratner:15a}
By
postulating a Lorentzian transmission,
an analytical formula
\cite{Stafford:96,Metzger:01b,Baldea:2010e,Baldea:2012b,Ratner:15a}
for the current $I$ as a function of the applied bias $V$ can easily be derived
\begin{equation*}
  I  = \frac{G_0}{e}
    \ \frac{2 \Gamma_{s} \Gamma_{t}}{\Gamma_{s} + \Gamma_{t}}
    \left(\arctan \frac{2 \varepsilon_{0} + e V}{\Gamma_{s} + \Gamma_{t}} -
    \arctan \frac{2 \varepsilon_{0} - e V}{\Gamma_{s} + \Gamma_{t}} \right)
\label{eq-I-arctan}
\end{equation*}
Here, $e$ is the elementary charge, $G_0 = e^2/h = 77.48\,\mu\mbox{S}$ is the conductance quantum,
and $\varepsilon_0 = E_{MO} - E_F$ the energy offset
relative to the Fermi energy ($E_F$). 
Inspection of this formula immediately reveals
that (in cases where $\varepsilon_0$
does not depend on $V$, see also \secname\ref{sec:remarks})
the $I$-$V$ curve is strictly symmetric irrespective whether the MO couplings to electrodes
$\Gamma_{s}$ and $\Gamma_{t}$ are equal or not. Rephrasing, $\Gamma_{s} \neq \Gamma_{t}$ does not result in
current rectification
\begin{equation*}
\mbox{RR}(V) \equiv -I(V)/I(-V) \neq 1 
\label{eq-RR}
\end{equation*}
The Lorentzian transmission is a phenomenological assumption that deserves
quantum mechanical justification, at least based on a reasonable model Hamiltonian.
Calculations using Keldysh' nonequilibrium formalism show that transmission is Lorentzian
if the embedding self-energies $\Sigma_{s,t}$ quantifying the MO coupling
to electrodes are assumed to be purely imaginary and energy independent
\begin{equation*}
\Sigma_{s,t} = -\frac{i}{2} \Gamma_{s,t}
\label{eq-Sigma_x-imag}
\end{equation*}
Still, even for a simple model Hamiltonian like that expressed by \gl~(\ref{eq-H}) below,
the embedding self-energies are neither purely imaginary nor energy independent (see \secname\ref{sec:formulas}).
Do deviations of $\Sigma_{s,t}$ from the above form 
make it possible that merely unequal couplings ($\Gamma_s \neq \Gamma_t$)
result in an observable current rectification ($\mbox{RR} \neq 1$)?

\ib{Demonstrating that this is not the case is the general aim of this paper. Drawing attention 
  on the incorrectness of a never-demonstrated formula (namely, \gl~(\ref{eq-CS}) below) yielding $\mbox{RR} \neq 1$ 
  for $\Gamma_{s} \neq \Gamma_{t}$ utilized in previous publications to quantitatively analyze
  current rectification in real molecular junctions is the most important specific aim of the present report.}
\section{The Single Level Model}
\label{sec:model}
Let us consider the steady-state charge transport in a two-terminal setup
consisting of a molecule (M) modeled as a single energy level (``molecular orbital'' MO) $\varepsilon_0$
linked to two electrodes referred to as ``substrate'' (label $s$) and ``tip'' (label $t$) subject to an external bias $V$.
A general second quantized full Hamiltonian
describing the charge transport mediated by a single energy level reads \cite{Baldea:2010e}
\begin{eqnarray}
H & = & 
\underbrace{\sum_{l \leq -1} \left[\mu_s c_l^{\dagger} c_l - \left( t_s c_{l}^{\dagger} c_{l-1} + H.c.\right)\right]}_{H_{s}}
+
\underbrace{\sum_{r \geq 1} \left[\mu_t c_r^{\dagger} c_r - \left( t_t c_{r}^{\dagger} c_{r+1} + H.c.\right)\right]}_{H_{t}} \nonumber \\ 
&  & 
\underbrace{- \left(\tau_s c_{-1}^{\dagger} c_0 + \mbox{H.c.}\right)}_{H_{s,M}}
+ \underbrace{\varepsilon_0 c_0^{\dagger} c_0}_{H_M}
\ \underbrace{- \left(\tau_t c_{1}^{\dagger} c + \mbox{H.c.}\right)}_{H_{t,M}}
\label{eq-H} \\
\varepsilon_0 & \equiv & \varepsilon_0 (\ldots) \nonumber
\end{eqnarray}
Above, creation $c_{l,r}^\dagger$ and annihilation $c_{l,r}$
operators refer to single electron states in substrate's and tip's conduction band
of widths $W_{s,t} = 4 \left\vert t_x\right\vert$
respectively. The subscript $0$ refers to the single molecular level considered,
and $\tau_{s,t}$ are effective (average) exchange integrals quantifying the MO-electrode charge transfers.
For the sake of simplicity, electron spin will not be included explicitly
but its contribution (a multiplicative factor
of two) will be accounted for whenever physically relevant (e.g., \gl~(\ref{eq-I})).
Strong on-site Coulomb repulsion \cite{Stafford:96,Buttiker:03}
implicitly assumed in \gl~(\ref{eq-I}) precludes level double occupancy
and leaves Coulomb blockade and Kondo effect beyond present consideration.
(Point to remember: typical ionization energies ($IP \sim 10$\,eV)
in real molecules are very large, much larger than
(HO)MO offsets relative to electrode's Fermi energy $\left\vert \varepsilon_0 \right\vert \alt 1$\,eV.)

In the presently considered zero temperature case, single-particle electron states
in electrodes are filled up to energies
below 
the electrodes' (electro)chemical potential $\mu_{x}$ ($x=s,t$) whose imbalance
\begin{equation}
\label{eq-mu}
\mu_{s,t} = E_F \pm e V/2;\ \mu_s - \mu_t = e V
\end{equation}
caused by an applied bias $V$ gives raise to an electric current through junction
on which we will focus next. Notice that by virtue of \gl~(\ref{eq-mu}) $V>0$ means
a positive t(ip) electrode, an aspect of practical relevance when discussing
the direction of rectification in a specific real junctions.

Noteworthily, \gl~(\ref{eq-I}) does by no means rule out a 
$V$-dependence of $\varepsilon_0$. This may arise when the level (MO) center-of-charge
is located asymmetrically with respect to electrodes
(``lever'' \cite{Metzger:01b} or ``potentiometer'' \cite{Baldea:2015g,Baldea:2018c} rule)
or due to intramolecular Stark effect,
a point to which we will return in the end of \secname\ref{sec:remarks}.
It is what we meant while writing ``$\ldots$'' in the last line there.

\ib{Provided that the Hamiltonian is the same for ``forward'' and ``backward'' current flow, there cannot be rectification.
This is a completely general result, independent on details of models; an example is
what was called zero current theorem in studies on generic tight binding models.\cite{Todorov:93,Todorov:02}
The point with the specific case presently considered is that, albeit oversimplified,
provided that $t_{s} \neq t_{t}$ and $\tau_{s} \neq \tau_{t}$ 
the model Hamiltonian of \gl~(\ref{eq-H}) 
does allow forward and backward currents be different $I(-V) \neq -I(V)$
even if $\varepsilon_0$ does not depend on $V$.\label{page:zero-current} So, in principle
\gl~(\ref{eq-H}) is compatible with $ RR(V) \neq 1$. This is substantiated
by \gl~(\ref{eq-renormalized-e0}) deduced below; an exact result, demonstrating that
current rectification can occur. The important problem is, however,
whether this broken ``left-right'' symmetry
($c_{l} \rightleftharpoons c_{r}, c_{l}^{\dagger} \rightleftharpoons c_{r}^{\dagger}, V \rightleftharpoons -V$)
of \gl~(\ref{eq-H}) can be source of an \emph{observable} current rectification.}

\section{General Results}
\label{sec:formulas}
Within the Keldysh formalism,\cite{mahan} the key quantity needed to express
the current $I$ (see \gl~(\ref{eq-I}) below \cite{Caroli:71a,Meir:94,HaugJauho,Xue:01}) through a molecular junction 
is the retarded Green's function $G^{R}$ of the ``embedded'' molecule.
It is related to the retarded Green's function of the isolated molecule
\begin{equation*}
\label{eq-G_0}
G_{0}^{R}(\varepsilon) = 1/\left( \varepsilon - \varepsilon_0 + i 0^{+}\right)
\end{equation*}
via Dyson's equation
\begin{equation}
\label{eq-dyson}
\left[ G^{R}(\varepsilon)\right]^{-1} =
\left[ G_{0}^{R}(\varepsilon)\right]^{-1} - \Sigma_{s}(\varepsilon) -  \Sigma_{t}(\varepsilon)
\end{equation}
The embedding self-energies $\Sigma_{s,t}$ have the form \cite{Meir:92,HaugJauho,Datta:05}
\begin{equation}
\label{eq-Sigma_x}
\Sigma_{x}(\varepsilon) \equiv \Delta_{x}(\varepsilon) - \frac{i}{2} \Gamma_{x}(\varepsilon)
= \vert \tau_{x}\vert^2 g_{x}^{R}(\varepsilon) 
\end{equation}
and account for the MO coupling to electrodes via the average exchange integrals of the MO-electrode
couplings $\tau_x$.\cite{Meir:92,Meir:94} 
The (surface) retarded Green's functions
of the semi-infinite electrodes can be expressed in closed analytical forms\cite{Newns:69b,desjonqueres:96,Datta:05,Peskin:10}
\begin{equation}
\label{eq-g_x}
g_{x}^{R}(\varepsilon) = 8 \frac{\varepsilon - \mu_x}{W_{x}^2} -
i \frac{4}{W_x} \sqrt{1 - 4 \left(\frac{\varepsilon - \mu_x}{W_x}\right)^2}
\end{equation}
Below, we confine ourselves to typical situations where applied biases yield
imbalance of electrodes' electrochemical potential sufficiently smaller
that electrodes' bandwidth 
$e \vert V \vert < W_x / 2 $, ensuring thereby
that the square root entering the
RHS of \gls~(\ref{eq-g_x}) and (\ref{eq-Gamma_x-ex}) are real numbers.
Otherwise, electrodes' finite band may give rise to negative differential resistance effects,
as discussed elsewhere.\cite{Baldea:2010e}

The electrode's density of states (DOS) $\rho_{x}(\varepsilon)$ can be written as
\begin{equation}
\label{eq-rho_x}
\rho_x(\varepsilon) \equiv - \frac{1}{\pi} \mbox{\, Im\,} g_{x}^{R}(\varepsilon) = \frac{4}{\pi}\frac{1}{W_x} \sqrt{1 - 4 \left(\frac{\varepsilon - \mu_x}{W_x}\right)^2}
\end{equation}
and has at the Fermi energy a value
\begin{equation}
\label{eq-rho_x-F}
\rho_x \equiv \left . \rho_x(\varepsilon)\right\vert_{\varepsilon = \mu_x} = \frac{4}{\pi} \frac{1}{W_x} = \frac{1.2733}{W_x}
\end{equation}
which is basically the inverse of electrode's conduction bandwidth.

Based on the aforementioned, analytical forms for $\Delta_x$ and $\Gamma_x$ can be deduced
\cite{Baldea:2010e} 
\begin{subequations}
\label{eq-Gamma_x-Delta_x-ex}
\begin{eqnarray}
\label{eq-Gamma_x-ex}
\Gamma_{x}(\varepsilon) & = &
\Gamma_x \sqrt{1 -
4 \left(
\frac{\varepsilon - \mu_x} {W_x}
\right)^2 }; \ \Gamma_x \equiv 8 \frac{\left\vert \tau_x\right\vert^2}{W_x} 
\\
\label{eq-Delta_x-ex} 
\Delta_{x}(\varepsilon) & = &
\frac{\Gamma_x}{W_x} \left(\varepsilon - \mu_x\right) =
\frac{\pi}{4} \Gamma_{x} \rho_{x} \left(\varepsilon - \mu_x\right) 
\end{eqnarray}
\end{subequations}
One should note here that the expression of $\Gamma_x$ in terms of $\rho_x$
deduced from \gls~(\ref{eq-rho_x-F}) and (\ref{eq-Gamma_x-ex}) 
\begin{equation*}
\label{eq-Gamma_x-0}
\Gamma_x = 2 \pi \rho_{x} \left\vert \tau_x\right\vert^2
\end{equation*}
is not restricted to the case of semi-elliptic DOS
of \gl~(\ref{eq-rho_x}).\cite{Schmickler:86,Meir:94,Yaliraki:98a,Schmickler:03}
Including MO-electrodes interactions beyond the choice for $H_{x,M}$ adopted in \gl~(\ref{eq-H}) 
is possible,\cite{Yaliraki:98a,Reuter:10} 
but will not be attempted here
because pertaining corrections were shown \cite{Hush:00} not to
substantially alter conclusions based on \gl~(\ref{eq-rho_x}). 

The closed form of the retarded Green's function describing the junction under applied bias ($V \neq 0$)
can now be obtained by inserting \gls~(\ref{eq-Gamma_x-ex}) and (\ref{eq-Delta_x-ex}) into the Dyson
equation (\ref{eq-dyson}) 
\begin{subequations}
\begin{equation}
\label{eq-renormalized-G_R}
G^R(\varepsilon) =
\frac{1}{1 - \frac{\Gamma_s}{W_s} - \frac{\Gamma_t}{W_t} }  
\frac{1}
{\varepsilon
- \tilde{\varepsilon}_{0}(V)
+ \frac{i}{2}\left[\tilde{\Gamma}_{s}(\varepsilon) + \tilde{\Gamma}_{t}(\varepsilon)\right]
}
\end{equation}
\begin{equation}
\label{eq-Gamma_s-ren}
\tilde{\Gamma}_{s}(\varepsilon) \equiv
    \Gamma_s \frac{\sqrt{1 - 4 \left(\frac{\varepsilon - eV/2}{W_s}\right)^2}}{1 - \frac{\Gamma_s}{W_s} - \frac{\Gamma_t}{W_t} } 
\end{equation}
\begin{equation}
\label{eq-Gamma_t-ren}
\tilde{\Gamma}_{t}(\varepsilon) \equiv
    \Gamma_t \frac{\sqrt{1 - 4 \left(\frac{\varepsilon + eV/2}{W_s}\right)^2}}{1 - \frac{\Gamma_s}{W_s} - \frac{\Gamma_t}{W_t} } 
\end{equation}
\end{subequations}
The retarded Green's function has ``pole''
(more precisely, this is the position where the real part of $\left[G^{R}(\varepsilon)\right]^{-1}$ vanishes) at 
\begin{subequations}
\label{eq-renormalized-e0}
\begin{eqnarray}
\label{eq-renormalized-e0-def}
\tilde{\varepsilon}_{0}(V) & = & \tilde{\varepsilon}_{0} + \delta \tilde{\varepsilon}_{0}(V) = \tilde{\varepsilon}_{0} + \gamma e V
\\
\label{eq-renormalized-e0-Delta}
\tilde{\varepsilon}_{0} & = & \frac{\varepsilon_0} {1 - \frac{\Gamma_s}{W_s} - \frac{\Gamma_t}{W_t}}
= \varepsilon_{0} \left(1 + \frac{\Gamma_s}{W_s} + \frac{\Gamma_t}{W_t}\right) + \mathcal{O}\left(\frac{\Gamma_{s,t}}{W_{s,t}}\right)^2
\\
\label{eq-renormalized-e0-gamma}
\gamma & = & \frac{1}{2}\frac{\frac{\Gamma_t}{W_t} - \frac{\Gamma_s}{W_s}} {1 - \frac{\Gamma_s}{W_s} - \frac{\Gamma_t}{W_t}} 
= \frac{1}{2}\left(\frac{\Gamma_t}{W_t} - \frac{\Gamma_s}{W_s}\right) + \mathcal{O}\left(\frac{\Gamma_{s,t}}{W_{s,t}}\right)^2
\end{eqnarray}
\end{subequations}
which defines MO energy $\tilde{\varepsilon}_{0}(V)$ of the embedded
molecule in a current carrying state.
Notice that \gl~(\ref{eq-renormalized-e0}) includes both the MO energy renormalization
due to the couplings to electrodes ($\Gamma_{s,t} \neq 0$) of the molecule embedded in the
unbiased ($V\equiv 0$) junction ($\varepsilon_0 \to \tilde{\varepsilon}_{0} \neq \varepsilon_0$)
and the bias driven MO energy renormalization ($V \neq 0 \to \delta \tilde{\varepsilon}_{0}(V) = \gamma e V \neq 0$).

Inserting the above expressions into the general formula \cite{Caroli:71a,Meir:92,HaugJauho}
\begin{equation}
\label{eq-I}
I = \frac{2 e}{h} \int_{\mu_s}^{\mu_t} d\,\varepsilon \Gamma_{s}(\varepsilon) \Gamma_{t}(\varepsilon)
\left\vert G^{R}(\varepsilon) \right\vert^2
\end{equation}
we are led to the general expression of the current determined by a single transport channel
(``single level model'') at zero temperature
\begin{equation}
\label{eq-I-specific}
  I = \frac{2 e}{h}
\int_{-eV/2}^{e V/2} \frac{\tilde{\Gamma}_{s}(\varepsilon) \tilde{\Gamma}_{t}(\varepsilon)}
       {\left[\varepsilon - \tilde{\varepsilon}_0(V)\right]^2 +
         \frac{\left[
         \tilde{\Gamma}_{s}(\varepsilon) + \tilde{\Gamma}_{t}(\varepsilon)\right]^2}{4}}
         d\,\varepsilon
\end{equation}

Along with \gl~(\ref{eq-I-specific}), the expression of the local density of states
\begin{equation}
\displaystyle
\label{eq-LDOS}
LDOS(\varepsilon) \equiv -\frac{1}{\pi} \mbox{Im} G^{R}(\varepsilon) =
\frac{1}{\pi}\frac{1}{1 - \frac{\Gamma_s}{W_s} - \frac{\Gamma_t}{W_t}}
\frac{\frac{1}{2}\left[\tilde{\Gamma}_{s}(\varepsilon) + \tilde{\Gamma}_{t}(\varepsilon)\right]}{\left[\varepsilon - \tilde{\varepsilon}_{0}(V)\right]^2 
   + \frac{1}{4}\left[ \tilde{\Gamma}_{s}(\varepsilon) + \tilde{\Gamma}_{t}(\varepsilon)\right]^2}
\end{equation}
better allows to emphasize the twofold role played by $\tilde{\Gamma}_{s,t}(\varepsilon)$:
renormalized MO couplings to electrodes (entering as multiplicative factors in \gl~(\ref{eq-I-specific})) and
renormalized partial level broadenings (cf.~\gl~(\ref{eq-LDOS})).

\Gl~(\ref{eq-renormalized-e0}) allows us to disentangle the impact of the
MO couplings' renormalization ($\Gamma_{s,t} \leftarrow \tilde{\Gamma}_{s,t}\left(\varepsilon\right), I_{\Gamma} \leftarrow I$)
\begin{equation}
\label{eq-I-Gamma}
  I_{\Gamma} = \frac{2 e}{h}
\int_{-eV/2}^{e V/2} \frac{\tilde{\Gamma}_{s}(\varepsilon) \tilde{\Gamma}_{t}(\varepsilon)}
       {\left(\varepsilon - \varepsilon_0\right)^2 +
         \frac{\left[
         \tilde{\Gamma}_{s}(\varepsilon) + \tilde{\Gamma}_{t}(\varepsilon)\right]^2}{4}}
         d\,\varepsilon
\end{equation}
from the impact of the MO energy renormalization ($\varepsilon_0 \leftarrow \tilde{\varepsilon}_{0}(V), I_{\varepsilon} \leftarrow I$)
\begin{equation*}
\label{eq-I-epsilon}
  I_{\varepsilon} = \frac{2 e}{h}
\frac{1}{\left(1 - \frac{\Gamma_s}{W_s} - \frac{\Gamma_t}{W_t}\right)^2 } 
\int_{-eV/2}^{e V/2} \frac{\Gamma_{s} \Gamma_{t} }
       {\left[\varepsilon - \tilde{\varepsilon}_0(V)\right]^2 +
         \frac{\left(\Gamma_{s} + \Gamma_{t}
         \right)^2}{4}}
         d\,\varepsilon
\end{equation*}
The latter can be integrated out in closed form and reads
\begin{eqnarray}
  I_{\varepsilon}  & = & \frac{G_0/e}{\left(1 - \frac{\Gamma_s}{W_s} - \frac{\Gamma_t}{W_t}\right)^2 }  
    \ \frac{2 \Gamma_{s} \Gamma_{t}}{\Gamma_{s} + \Gamma_{t} 
    } \times \nonumber \\
    & \times & \left[\arctan \frac{2 \tilde{\varepsilon}_{0}(V) + e V}{\Gamma_{s} + \Gamma_{t} 
    } - \arctan \frac{2 \tilde{\varepsilon}_{0}(V) - e V}{\Gamma_{s} + \Gamma_{t} 
    } \right]
\label{eq-I-specific-approx-arctan}
\end{eqnarray}
where $\tilde{\varepsilon}_{0}(V)$ is given by \gl~(\ref{eq-renormalized-e0}).
If the charge transport occurs sufficiently far away from resonance (which is the usual case
\cite{Baldea:2019d,Baldea:2019h}), i.e.
\begin{equation*}
\label{eq-off-resonance}
   \left[2 \left\vert \tilde{\varepsilon}_{0}(V) - \right\vert e V\vert \right] /
   \left[ \Gamma_{s} + \Gamma_{t} 
   \right] \gg 1
\end{equation*}
\gl~(\ref{eq-I-specific-approx-arctan}) is amenable at the simpler form
\cite{Baldea:2012a}
\begin{equation}
\label{eq-I-ib}
  I_{\varepsilon} \simeq I_{\mbox{off-res}} = \frac{\Gamma_{s} \Gamma_{t}}{\left(1 - \frac{\Gamma_s}{W_s} - \frac{\Gamma_t}{W_t}\right)^2 }
    \ \frac{G_{0} V }{\tilde{\varepsilon}_{0}^{2}(V) - (e V/2)^2}
    \end{equation}
In the wide-band limit ($W_{s,t} \to \infty$),
$\Gamma_{s,t}(\varepsilon) \to \Gamma_{s,t}$, $\tilde{\varepsilon}_{0}(V) \to \varepsilon_0$,
and \gls~(\ref{eq-I-specific-approx-arctan}) and (\ref{eq-I-ib}) reduce to \gls~(3) and (4)
of ref.~\citenum{Baldea:2012a}. 

As a hybrid approximation, one can also consider couplings' renormalization only
in the numerator of the integrand entering the RHS of \gl~(\ref{eq-I-specific})
\begin{equation}
\label{eq-I-Gamma-partial}
  I_{\varepsilon (\Gamma)} = \frac{2 e}{h}
\int_{-eV/2}^{e V/2} \frac{
 \tilde{\Gamma}_{s}(\varepsilon) \tilde{\Gamma}_{t}(\varepsilon)
 }  
       {\left[\varepsilon - \tilde{\varepsilon}_{0}(V)\right]^2 +
         \frac{\left(
         \Gamma_{s} + \Gamma_{t} 
\right)^2
}{4}}
         d\,\varepsilon
\end{equation}
\section{Discussion}
\label{sec:disc}
Inspection of \Gl~(\ref{eq-I-specific}) reveals that, \emph{in principle},
rectification $I(-V) \neq - I(V)$ could arise in junctions whose electrodes have different properties
($\Gamma_s \neq \Gamma_t, W_s \neq W_t$).
Based on it, we will next interrogate how \emph{effective} the impact of electrodes' asymmetry on
current rectification in real junctions is.
\subsection{General Considerations}
\label{sec:disc-general}
According to \gl~(\ref{eq-I-specific}) $RR \neq 1$ (i) could be the result of a polarity dependent
bias-driven MO shift due to couplings' asymmetry,
$\delta \tilde{\varepsilon}_{0}(-V) \neq \delta \tilde{\varepsilon}_{0}(V)$
and/or because the expressions the parentheses under the square root entering the
\gls~(\ref{eq-Gamma_s-ren}) and (\ref{eq-Gamma_t-ren})
pertaining to the substrate and tip electrodes 
are (ii) significantly different of each other
and at least one of them is significantly different from zero.

To (i):
\Gl~(\ref{eq-renormalized-e0-gamma}) yields
\begin{equation*}
\label{eq-renormalized-e0-gamma-O}
\gamma = 
\frac{1}{2}\left(\frac{\Gamma_t}{W_t} - \frac{\Gamma_s}{W_s}\right) + \mathcal{O}\left(\frac{\Gamma_{s,t}}{W_{s,t}}\right)^2
\approx \frac{1}{2}\left(\frac{\Gamma_t}{W_t} - \frac{\Gamma_s}{W_s}\right)
\end{equation*}
which should make it clear that a bias-driven MO energy shift can safely be ruled out.
Indeed, even if the effective value $W = 35.8$\,eV deduced for gold from the DOS $\rho = 0.035\,\mbox{eV}^{-1}$ \cite{Datta:97a,Hush:00}
via \gl~(\ref{eq-rho_x-F}) might be somewhat overestimated, it still 
substantiate the conclusion that $W$ is much larger than MO-electrode couplings $\Gamma$;
values of $\Gamma$ estimated for real junctions are
at most $\sim 10^{-1}$\,eV, typically a few meV.\cite{Baldea:2015d,Baldea:2019d,Baldea:2019h,Frisbie:21}

Although not directly related to rectification, we note in passing that, for the same reason,
a substantial change in MO energy offset \emph{merely} due to molecule embedding
(i.e., $\tilde{\varepsilon}_0 \neq \varepsilon_0$, cf.~\gl~(\ref{eq-renormalized-e0-Delta}))
cannot occur
\begin{equation*}
\label{eq-renormalized-e0-Delta-O}
\tilde{\varepsilon}_{0} 
= \varepsilon_{0} \left(1 + \frac{\Gamma_s}{W_s} + \frac{\Gamma_t}{W_t}\right) + \mathcal{O}\left(\frac{\Gamma_{s,t}}{W_{s,t}}\right)^2
\approx \varepsilon_{0} \left(1 + \frac{\Gamma_s}{W_s} + \frac{\Gamma_t}{W_t}\right) \approx \varepsilon_0
\end{equation*}

To (ii):
Given the fact that the integration variable entering the RHS of \gl~(\ref{eq-I-specific})
varies in the range $ \left\vert \varepsilon \right\vert < e \vert V\vert /2$,
the maximum value of the parentheses under the square root are $\left(V/W_{s,t}\right)^2$.
This again shows that, at the highest bias values $V \sim 1 $\,V applied in real experiments,
differences between currents at positive and negative polarities can hardly exceed $\sim 0.1$\%.
\subsection{Specific Examples: Two Benchmark Cases}
\label{sec:c8t-opt3}
Having said this in general, let us focus on two benchmark junctions
fabricated with octanethiol (C8T) and 1, 1', 4', 1''-terphenyl-4-thiol (OPT3
in ref.~\citenum{Baldea:2019h}) molecules.
The parameters $\varepsilon_0 \equiv - \varepsilon_h$, $\Gamma_s$, and $\Gamma_t$ 
that make this analysis possible are available or can be estimated
thanks to recent extensive
investigations on these monothiolates \cite{Baldea:2019d,Baldea:2019h} as well as on their
dithiolate (C8DT \cite{Baldea:2019h} and OPD3 \cite{Baldea:2015d}) counterparts.
Data for dithiols ($d$) are also needed because, while providing values of the (geometric)
average $\Gamma = \sqrt{\Gamma_s \Gamma_t}$,\cite{Baldea:2012a} transport data for a given molecular species
do not allow the separate determination of the two individual 
components $\Gamma_s$ and $\Gamma_t$ for the presently considered monothiols ($m$).

In view of the fact that not only SAMs deposited on gold but also junctions fabricated with those dithiols
are characterized by extremely small statistical variations in their transport properties,\cite{Baldea:2017e}
it is legitimate
to assume $\Gamma_{s}^{d} \approx \Gamma_{t}^{d} \approx \Gamma^{d}$;
dithiolate species form stable covalent bonds responsible for chemisorption
both at the substrate and at the tip.
In addition, one can assume
$\Gamma_{s}^{m} \approx \Gamma_{s}^{d} \approx \Gamma^{d}$; 
both monothiols and dithiols are linked to substrate by thiol groups.
Doing so, based on $\Gamma^{C8DT} = 14.88$\,meV \cite{Baldea:2019h} and $\Gamma^{C8T} = 2.45$\,meV \cite{Baldea:2019h}
we get  $\Gamma_{s}^{C8T} = 14.88$\,meV and $\Gamma_{t}^{C8T} = \left(\Gamma^{C8T}\right)^2/\Gamma_{s}^{C8T} = 0.40$\,meV.
Similarly, using  $\Gamma^{OPD3} = 18.34$\,meV \cite{Baldea:2019d} and $\Gamma^{OPT3} = 4.52$\,meV \cite{Baldea:2019d}
we estimate $\Gamma_{s}^{OPT3} = 18.34$\,meV and $\Gamma_{t}^{OPT3} = \left(\Gamma^{OPT3}\right)^2/\Gamma_{s}^{OPT3} = 1.11$\,meV.

\begin{figure*} 
  \centerline{
    \includegraphics[width=0.30\textwidth,height=0.20\textwidth,angle=0]{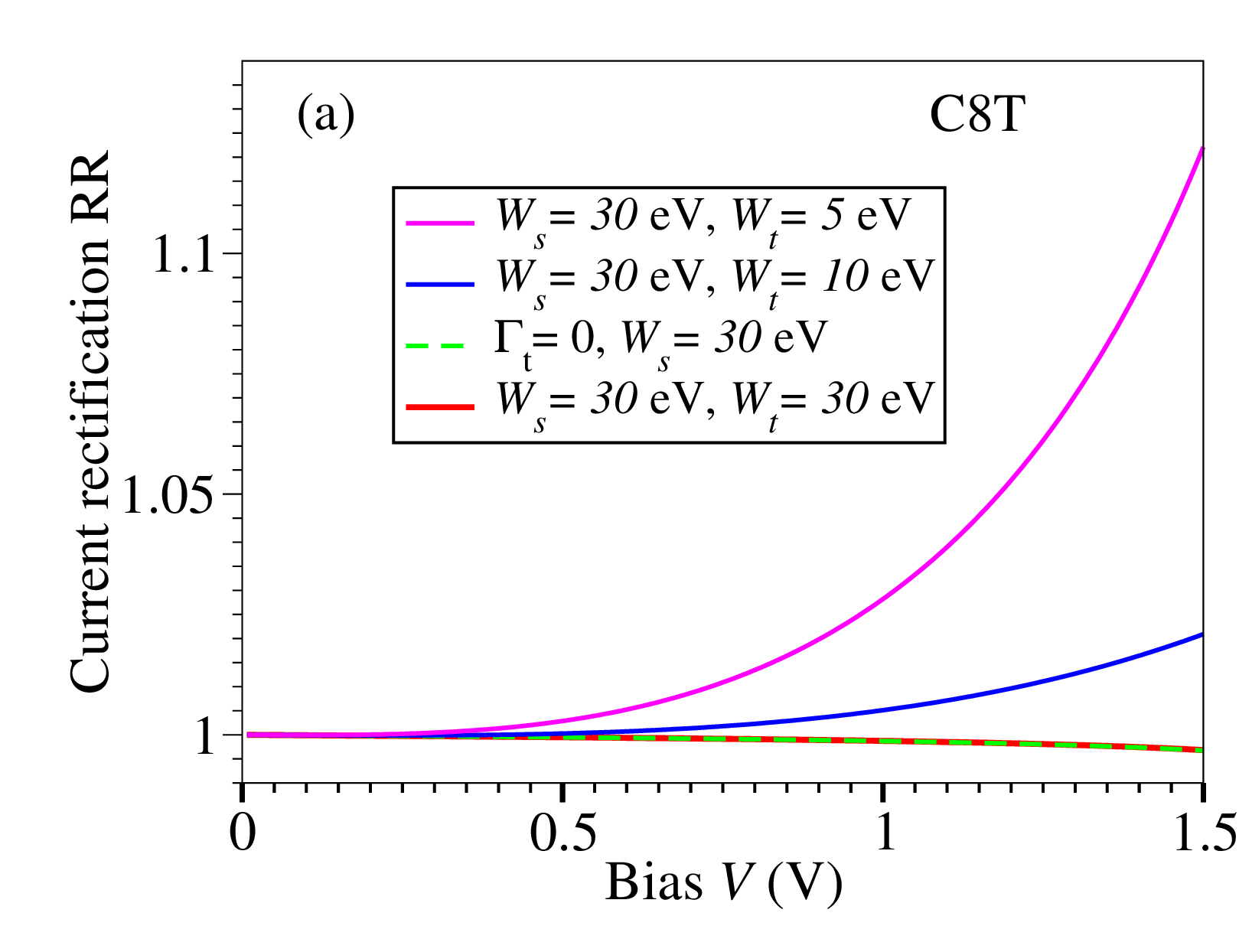}
    \includegraphics[width=0.30\textwidth,height=0.20\textwidth,angle=0]{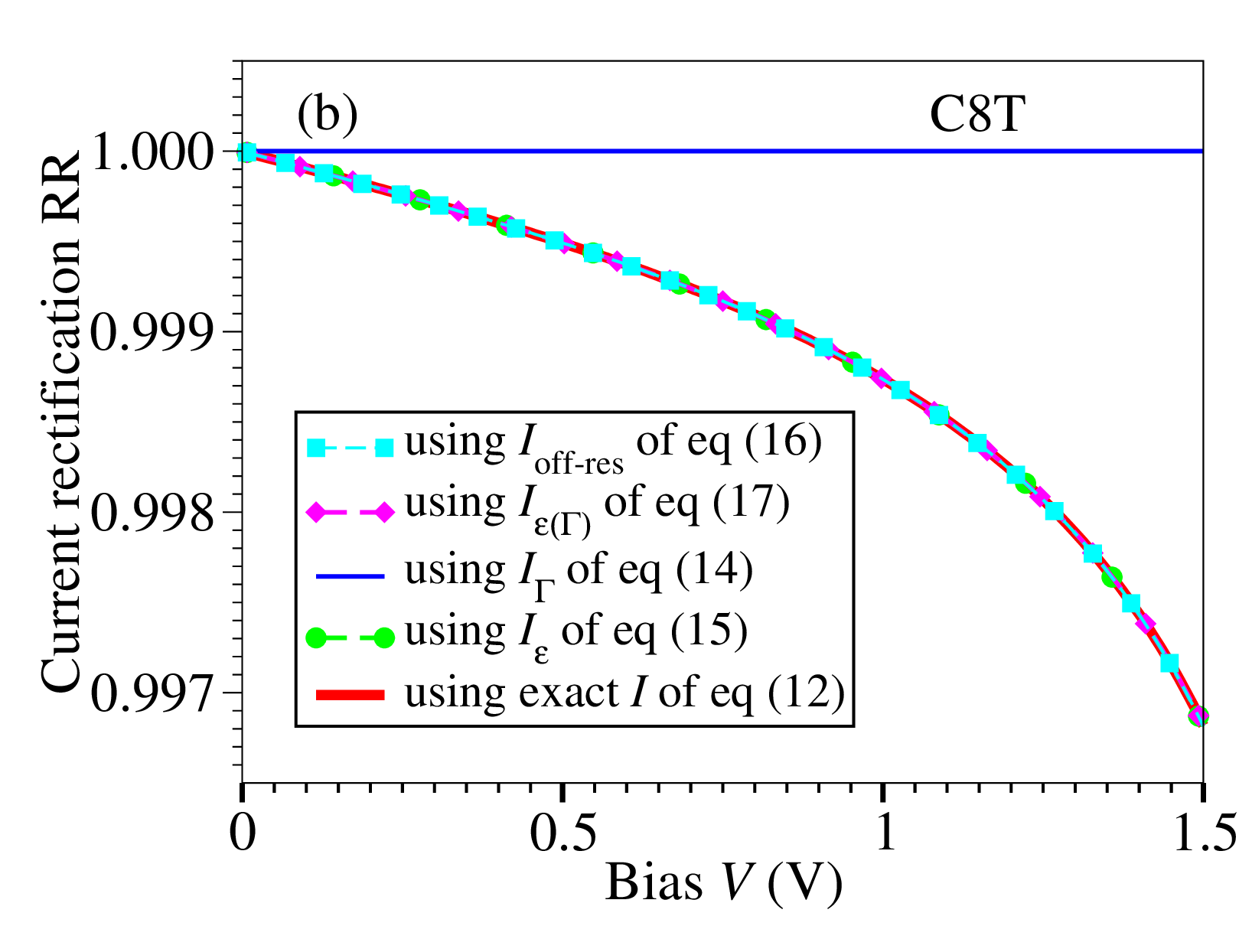}
  }
  \caption{Current rectification RR calculated in experimentally relevant bias range
  using parameters (a) estimated for C8T junctions \cite{Baldea:2019h} and (b) modified to overestimate
  the RR-values. See the main text in \secname\ref{sec:c8t-opt3} for details.}
  \label{fig:c8t}
\end{figure*}

\begin{figure*} 
  \centerline{
    \includegraphics[width=0.30\textwidth,height=0.20\textwidth,angle=0]{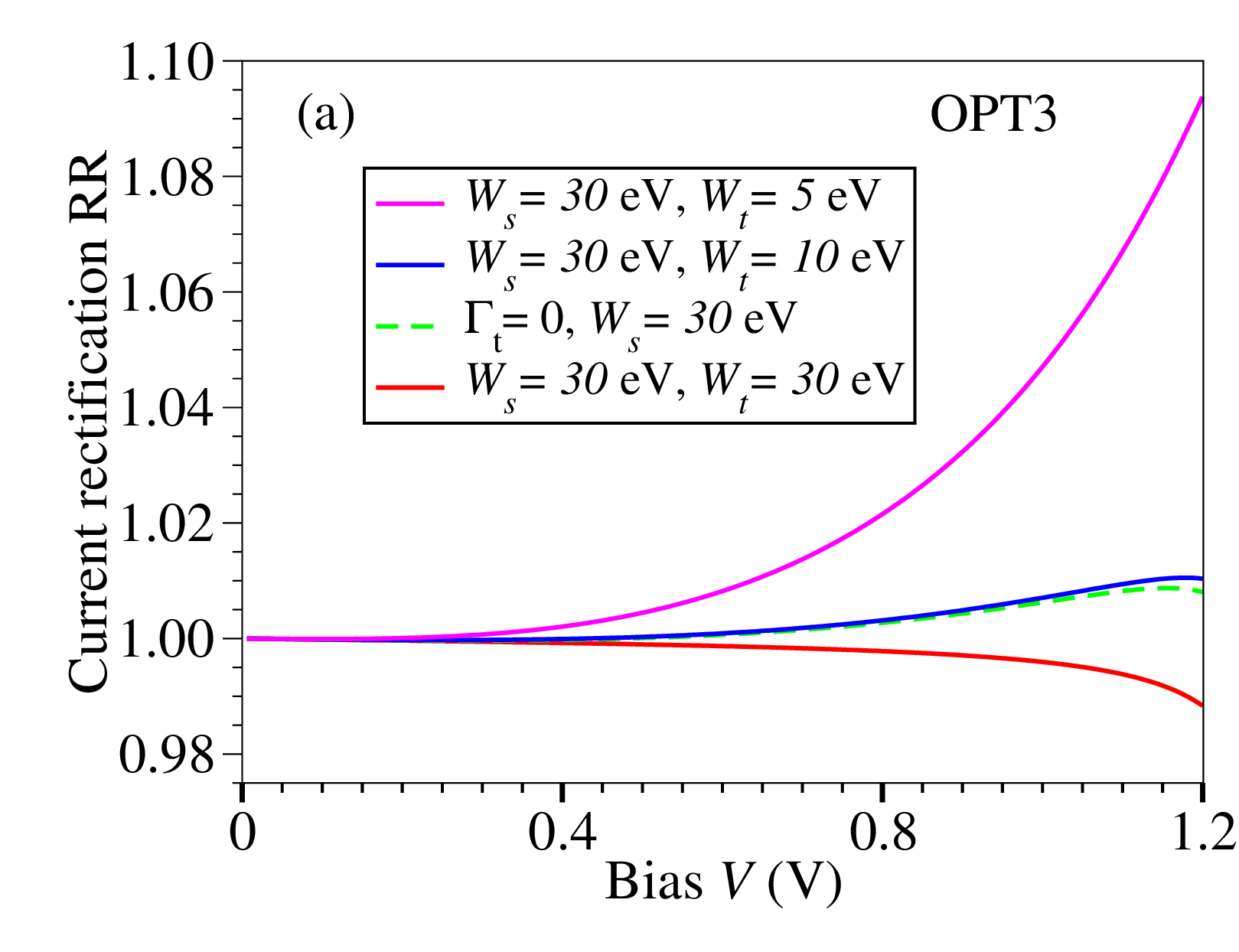}
    \includegraphics[width=0.30\textwidth,height=0.20\textwidth,angle=0]{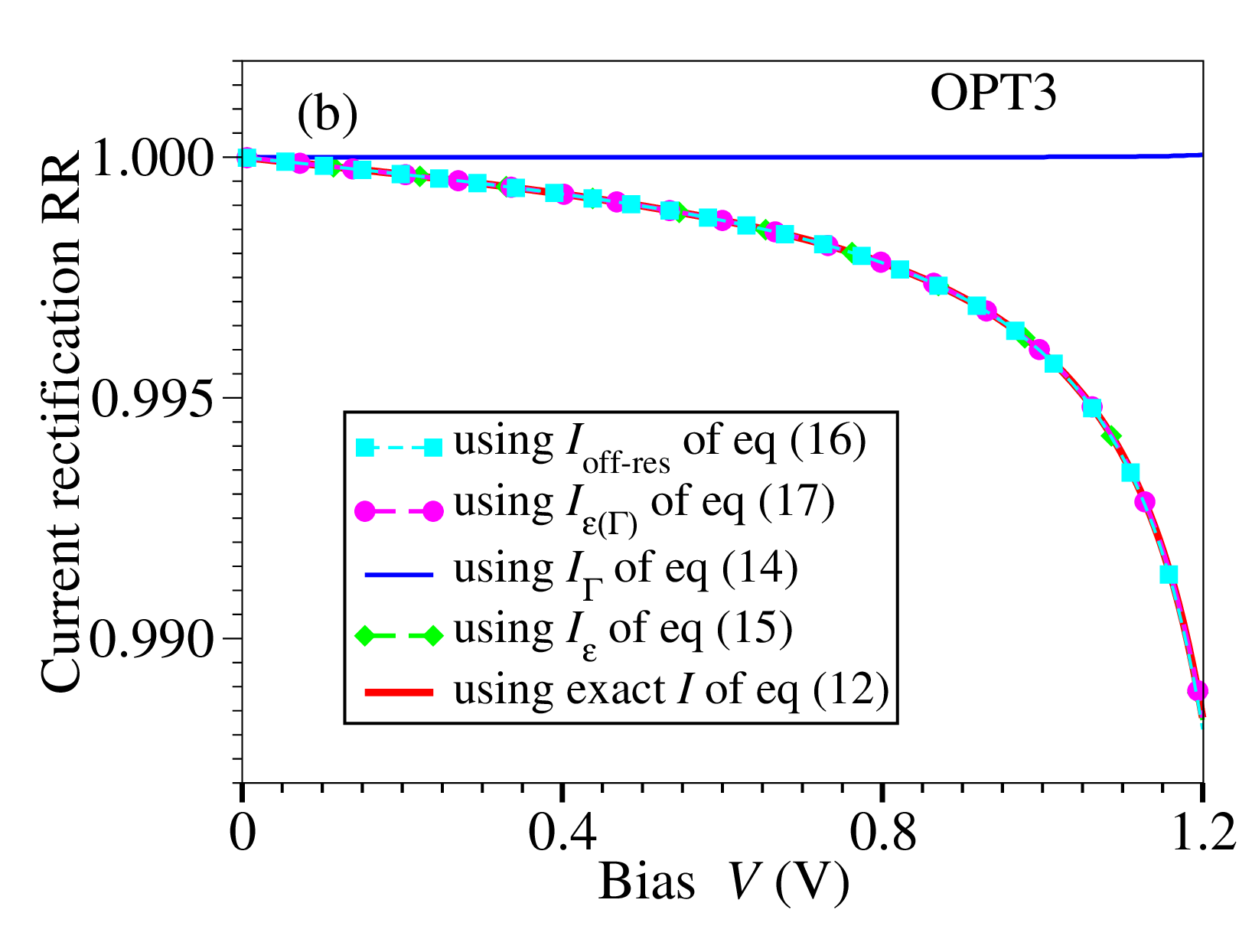}
  }
  \caption{Current rectification RR calculated in experimentally relevant bias range
  using parameters (a) estimated for OPT3 junctions \cite{Baldea:2019d} and (b) modified to overestimate
  the RR-values. See the main text in \secname\ref{sec:c8t-opt3} for details.}
  \label{fig:opt3}
\end{figure*}

\figsname\ref{fig:c8t} and \ref{fig:opt3} depict the dependence on bias of the current rectification
obtained by using the HOMO offsets derived from recent transport measurements 
($\varepsilon_{0}^{C8T} \equiv - \varepsilon_{h}^{C8T} = -1.01$\,eV \cite{Baldea:2019h}
and $\varepsilon_{0}^{OPT3} \equiv - \varepsilon_{h}^{OPT3} = -0.66$\,eV \cite{Baldea:2019d})
along with the aforementioned values of $\Gamma_{s,t}$.
As shown by the blue curves of \figsname\ref{fig:c8t}b and \ref{fig:opt3}b, the
impact of $\Gamma_{s,t}$-renormalization brought about by applied bias is completely negligible.
The ``largest'' contribution to rectification comes from the
renormalization of the HOMO energy, which is accounted for by \gl~(\ref{eq-I-epsilon}) and
depicted by the magenta curves. In off-resonant situations and biases of experimental interest
this effect is very accurately described by the simpler \gl~(\ref{eq-I-ib}), which
represents the generalization beyond the wide-band approximation of a result
(\gl~(4) of ref.~\citenum{Baldea:2012a})
deduced earlier in the limit $W_x \to \infty$.

To sum up, \figsname\ref{fig:c8t} and \ref{fig:opt3} clearly reveal that, when fully accounted for,
renormalization effects due to MO-couplings to the electrodes
of C8T and OPT3 junctions are unable to make RR significantly different from unity
and are by no means responsible for the values observed in experiment
($\mbox{RR}_{OPT3} \simeq 2.5 \mbox{ at } V = 1.2$\,V and
$\mbox{RR}_{C8T} \simeq 0.7 \mbox{ at } V = 1.5$\,V).\cite{Baldea:2018c,Baldea:2019d,Baldea:2019h} 
To make this point clearer, in addition to calculations based on parameter values deduced from
experimental data (see above), we also performed companion simulations to artificially enhance
the impact of the aforementioned renormalization, ruling out that possible parameters' inaccuracy may
vitiate the conclusions presented below. E.g.:

(i) We considered the case of extreme asymmetric couplings to
electrodes ($\Gamma_t \to 0$). Calculations for this case
(green curves in  \figsname\ref{fig:c8t}a and \ref{fig:opt3}a) yield
values of RR that cannot be practically distinguished from unity.

(ii) We performed simulations by using electrode bandwidths $W_t$ substantially smaller than
that previously estimated ($W_{Au} = 35.8$\,eV,\cite{Datta:97a,Hush:00} see above).
Letting alone the comparative purpose, the rationale for this choice might be that,
unlike practically infinite substrates, more or less sharp
tips may have DOS ($\rho_x \approx 1/W_x$, cf.~eq~(\ref{eq-rho_x}))
different from the value for infinite metal.
Notice again that by choosing a smaller $W_t$-value, renormalization effects are
(artificially) overestimated: $W$'s enter the denominators of relevant formulas, e.g.~\gl~(\ref{eq-I-specific}).
Results of these calculations are depicted by the blue and magenta curves in \figsname\ref{fig:c8t}a and \ref{fig:opt3}a.
The emerging conclusion is the same; although overestimated, this rectification $\mbox{RR} = 1 \pm 0.0\ldots $ substantially departs
from that deduced from experiment and misses any practical relevance.
\ib{
\subsection{Interrogating Possible Charge Accumulation Effects at Contacts}
\label{sec:occupancies}
The exact results reported above in this section have substantiated what the
title already stated: the $\Gamma_{s,t}$ couplings' asymmetry does not have a quantitatively relevant
impact on current rectification. But, after all, this conclusion is based on a highly simplified model.
So, one may wonder whether the unequal charge transfer rates $\Gamma_s \neq \Gamma_t$ may
still significantly enhance the completely negligible current asymmetry $I(+V) \neq \vert I(-V) \vert$
via physical effects escaping \gl~(\ref{eq-H}). Being driven by unequal  $\Gamma_s \neq \Gamma_t$,
the effect to be next discussed belongs to this category.

Fabrication of a molecular junction necessarily implies a certain (possibly asymmetric) charge exchange
between the embedded molecule and electrodes which cannot be ignored even
within a single-electron description \cite{Todorov:93} like that
underlying \gl~(\ref{eq-H}); the level's energy \emph{is} renormalized (cf.~\gl~(\ref{eq-renormalized-e0}))
but this is a wispy effect. If significantly dependent on bias polarity,
extra electronic charge accumulated at ``interfaces'' (positions $l=-1$ and $r=1$ \gl~(\ref{eq-H}))
might be relevant in the context of rectification. If this was the case and the MO occupancy
significantly changed upon bias polarity reversal, one would have to consider the
associated electrostatic interactions (preferably treated within a many-body picture,
since this turned out to be feasible\cite{Baldea:2012b}) as a potential source of rectification.

With this in mind, we computed the bias-dependent MO occupancy
$ n_0 = \langle c_{0}^\dagger c_{0}\rangle$ as well as the occupancies
$n_{s} \equiv \left . \langle c_{l}^{\dagger} c_{l}\rangle\right\vert_{l=-1}$ and
$n_{t} \equiv \left . \langle c_{r}^{\dagger} c_{r}\rangle\right\vert_{r=1}$ at the contacts.
This is an easy task
because, in the absence of electron correlations, the nonequilibrium Keldysh lesser Green's functions ($\mathbf{G}^{<}$)
needed can be straightforwardly expressed in terms of the retarded Green's functions.
Being a rather marginal issue in this paper we skip the technical details;
all relevant information can be found in ref.~\onlinecite{Baldea:2016a}.
The results of these calculations using the model
parameters deduced in \secname\ref{sec:c8t-opt3} for C8T are collected in \figname\ref{sec:occupancies}.
\begin{figure*} 
  \centerline{
    \includegraphics[width=0.28\textwidth,height=0.20\textwidth,angle=0]{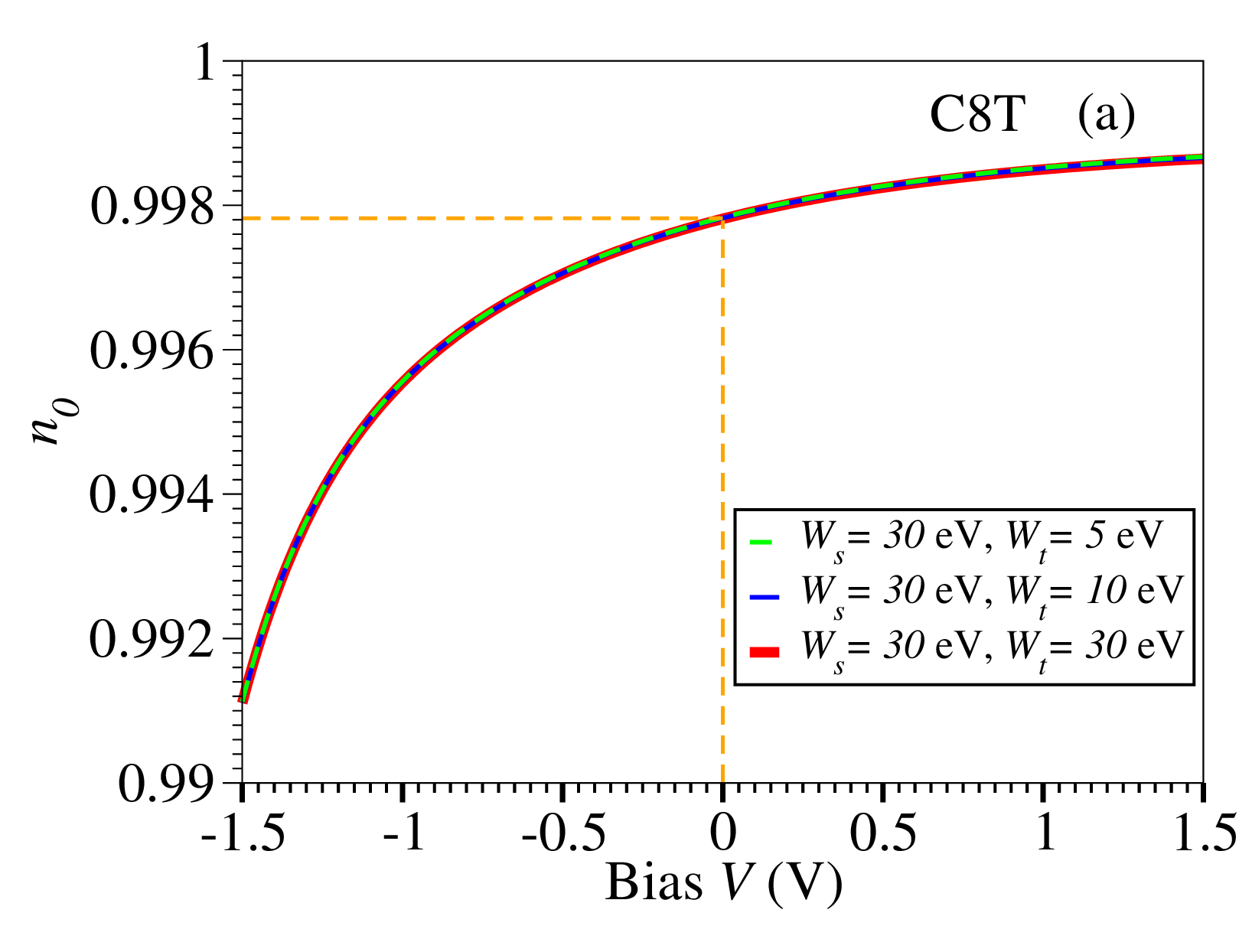}
    \includegraphics[width=0.28\textwidth,height=0.20\textwidth,angle=0]{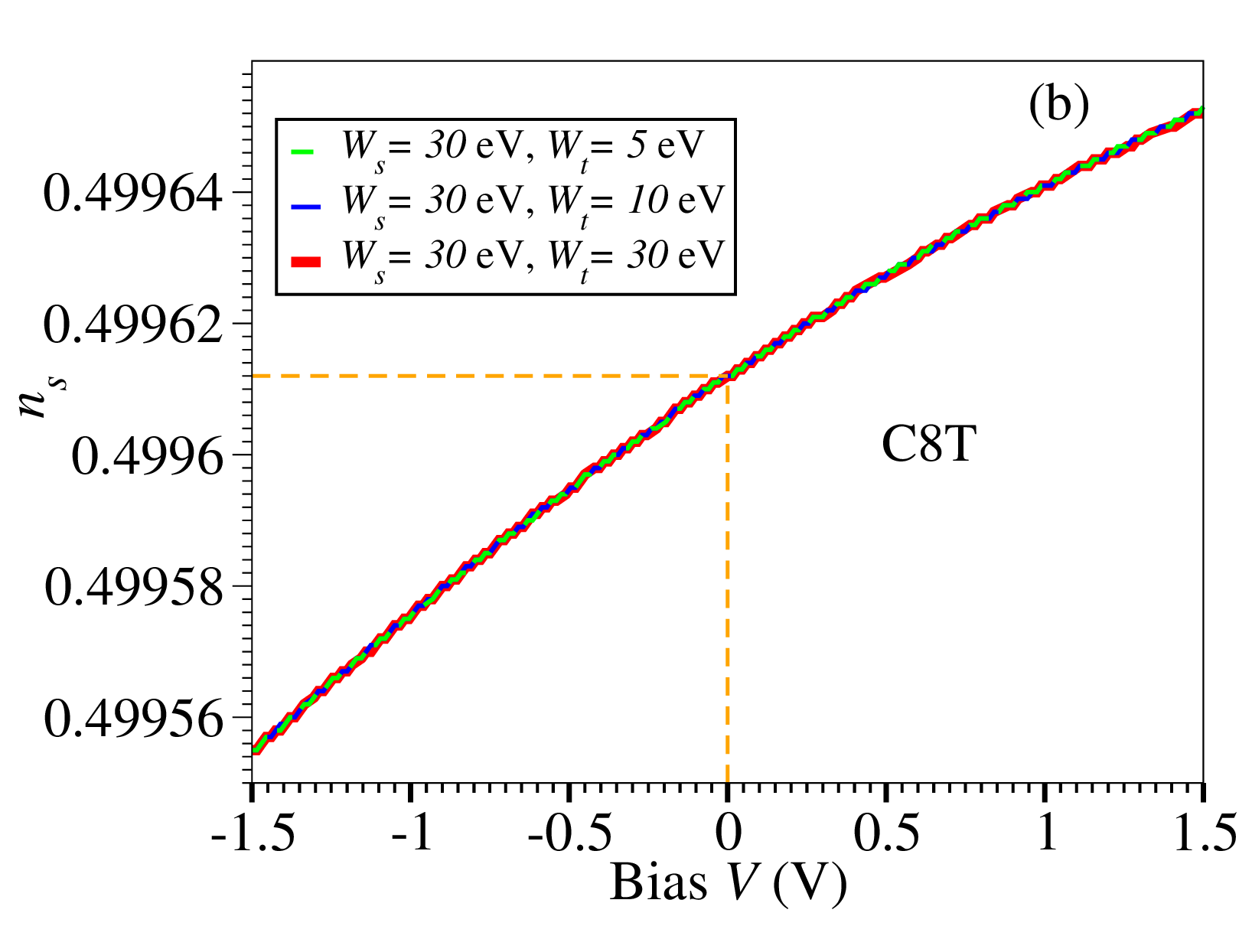}
    \includegraphics[width=0.28\textwidth,height=0.20\textwidth,angle=0]{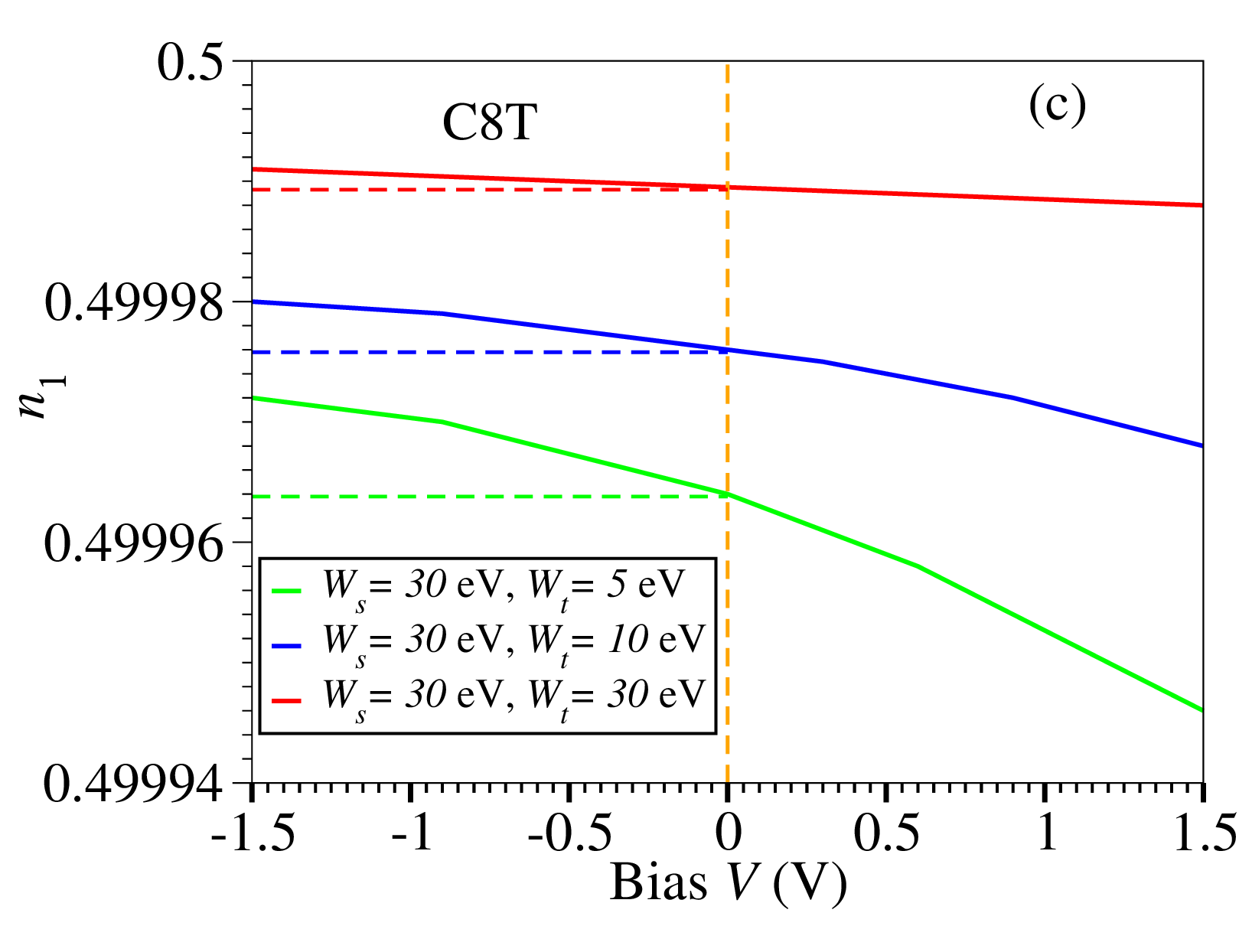}
  }
  \caption{The bias dependence of the occupancies $n_{j}$ of (a) the HOMO and the adjacent sites in (b) substrate and (c) tip electrodes
  computed using parameters estimated for C8T junctions \cite{Baldea:2019d} and modified to overestimate
  the RR-values. See \figname\ref{fig:c8t} and the main text (\secname\ref{sec:c8t-opt3}) for details.}
  \label{fig:occupancies}
\end{figure*}
Within the model considered, the differences $n_0 \neq 1$ and $n_{s,t} \neq 1/2$ visible in \figname\ref{sec:occupancies}
reflect the combined effect of coupling to electrodes ($\Gamma_{s,t} \neq 0$) and applied bias ($V\neq 0$).
The highly localized C8T's HOMO concentrated in the immediate vicinity of the substrate
(e.g, \figname~6 of ref~\onlinecite{Baldea:2015g}) makes the impact on $n_s$ ``stronger'' that that on $n_t$;
compare \figname\ref{sec:occupancies}b and \figname\ref{sec:occupancies}c among themselves.
Still, most importantly in the present context, for all biases of experimental relevance, 
$V$ has an altogether negligible impact on all electron occupancies. This rules out any
notable contribution to current rectification. 

Above, we intentionally restricted ourselves to the case of C8T.
The much less spatially asymmetric OPT3's HOMO delocalized over the entire molecule
(e.g., \figname~S7 of ref.~\onlinecite{Baldea:2019d})
makes the difference $n_s \neq n_t$ in this molecular species even substantially smaller than for C8T.
} 
\subsection{Additional Remarks}
\label{sec:remarks}
We do not want to end this work without commenting on 
earlier literature attempts to describe the current rectification
by postulating a bias-driven energy shift of the single dominant transport channel
depending on the coupling to electrodes $\Gamma_{s,t}$ \cite{CuevasScheer:10,Lee:11,Scheer:12,Nijhuis:15e}
as follows
\begin{subequations}
\label{eq-CS}
\begin{equation}
\label{eq-epsilon-CS}
V_s \equiv -\frac{V}{2} \leq V \leq V_t \equiv \frac{V}{2};
\left . \varepsilon_{0}\right\vert_{V=0} \xlongrightarrow{V\neq 0} \varepsilon_{0}(V) = \varepsilon_{0} + \overline{\gamma}\, e V
\end{equation}
\begin{equation}
\label{eq-gamma-CS}
\overline{\gamma} = \frac{1}{2}\frac{\Gamma_s - \Gamma_t}{\Gamma_s + \Gamma_t} = \frac{1}{2}(1 - \delta);\
\delta \equiv \frac{2 \Gamma_t}{\Gamma_s + \Gamma_t} 
\end{equation}
\end{subequations}
or equivalently
\begin{subequations}
\label{eq-CS-sym}
\begin{equation}
\label{eq-epsilon-CS-asym}
V_s^{\prime} \equiv 0 \leq V \leq V_t^{\prime} \equiv V;
\left . \varepsilon_{0}^{\prime}\right\vert_{V=0} \xlongrightarrow{V\neq 0} \varepsilon_{0}^{\prime}(V) =
\varepsilon_{0} 
+ \eta\, e V
\end{equation}
\begin{equation}
\label{eq-gamma-CS-asym}
\eta = \frac{\Gamma_s}{\Gamma_s + \Gamma_t} = \frac{1}{2} + \overline{\gamma} 
\end{equation}
\end{subequations}
Notice that due to the different choice the electric potential origin,
$\varepsilon_{0}^{\prime}(V) = \varepsilon_{0}(V) + e V/2$
and $\eta = 1/2 + \overline{\gamma}$.
$\mu_s - \mu_t = eV $ holds in both cases,
implying, e.g., a positive bias on the $t$(ip) for $V>0$.

Importantly for checking its validity against experimental data, \gl~(\ref{eq-CS})
predicts that the \emph{direction} of the MO bias-driven shift (upwards or downwards)
is merely dependent on the sign of the couplings' difference $\Gamma_{s} - \Gamma_{t}$
\begin{equation*}
\delta \varepsilon_{0}(V) \equiv \varepsilon_{0}(V) - \left . \varepsilon_{0}(V)\right\vert_{V=0} =
\overline{\gamma} e V = \frac{1}{2}\frac{\Gamma_s - \Gamma_t}{\Gamma_s + \Gamma_t} e V \propto \mbox{sign}(\Gamma_{s} - \Gamma_{t})\, \mbox{sign}V
\end{equation*}
which translates into a current rectification (RR) direction
(i.e., $\mbox{RR} >1$ or $\mbox{RR} < 1$) merely dependent on the sign of $\Gamma_{s} - \Gamma_{t}$
expressed as follows
\begin{equation*}
\mbox{RR}_{HOMO}(V>0) \equiv -\frac{I(V)}{I(-V)} \left\{
\begin{array}{ll}
> 1 & \mbox{ for } \overline{\gamma} > 0 \Rightarrow \Gamma_s > \Gamma_t \\
< 1 & \mbox{ for } \overline{\gamma} > 0 \Rightarrow \Gamma_s < \Gamma_t \\
\end{array}
\right . \mbox{ for HOMO-mediated conduction } (\varepsilon_0 < 0) 
\end{equation*}
and
\begin{equation*}
\mbox{RR}_{LUMO}(V>0) \equiv -\frac{I(V)}{I(-V)} \left\{
\begin{array}{ll}
< 1 & \mbox{ for } \overline{\gamma} > 0 \Rightarrow \Gamma_s > \Gamma_t \\
> 1 & \mbox{ for } \overline{\gamma} < 0 \Rightarrow \Gamma_s < \Gamma_t \\
\end{array}
\right . \mbox{ for LUMO-mediated conduction } (\varepsilon_0 > 0) 
\end{equation*}

Although neither deduced theoretically nor validated experimentally, $\overline{\gamma}$ of 
\gl~(\ref{eq-gamma-CS})
--- or the equivalent quantity $\eta = 1/2 + \overline{\gamma} = \Gamma_s /\left(\Gamma_s + \Gamma_t\right)$ mentioned above
---
was utilized in previous publications, e.g., for textbook, illustrative purposes \cite{CuevasScheer:10}
or (sometimes\cite{Lee:11})
aware of the fact that a $\Gamma_{s,t}$ asymmetry
similar to the asymmetry in the voltage drop is merely
an assumption made for the sake of simplicity.
The minor difference between $\eta = \Gamma_s /\left(\Gamma_s + \Gamma_t\right)$ \cite{CuevasScheer:10,Lee:11}
and $\overline{\gamma}$ of our \gl~(\ref{eq-gamma-CS}) is  
due to the different choice the potential origin; the former
chose $V_s = 0, V_t = V$ while we used $V_s = -V/2, V_t = V/2$, cf.~\gl~(\ref{eq-mu}).

As visible, the \emph{never} deduced \gl~(\ref{eq-gamma-CS}) has no resemblance with
our \gl~(\ref{eq-renormalized-e0-gamma}), a formula \emph{deduced} 
here within a general single level model.
Although of little practical importance because we already saw above that the bias-driven MO shift
due to $V \neq 0$ and coupling to electrodes expressed by \gl~(\ref{eq-renormalized-e0})
is altogether negligible, it could still be remarked that 
even the shift direction of the never-deduced \gl~(\ref{eq-CS}) may be problematic:
$\overline{\gamma} \propto \mbox{sign}\left(\Gamma_s - \Gamma_t\right)$ (cf.~\gl~(\ref{eq-gamma-CS}))
as opposed to $\gamma \propto \mbox{sign}\left(\Gamma_t - \Gamma_s\right)$
(cf.~\gl~(\ref{eq-renormalized-e0-gamma}) for $W_s = W_t$).

By and large, one should conclude
that \gl~(\ref{eq-CS}) has no theoretical support.
This analytical demonstration adds additional evidence to the fact emphasized earlier 
that $I$-$V$ asymmetry predicted by \gl~(\ref{eq-gamma-CS}) is at odds with
various experimental data collected under various platforms; see
refs.~\citenum{Baldea:2014f}, \citenum{Baldea:2015g}, and citations therein. 

\begin{figure*} 
  \centerline{
    \includegraphics[width=0.30\textwidth,height=0.20\textwidth,angle=0]{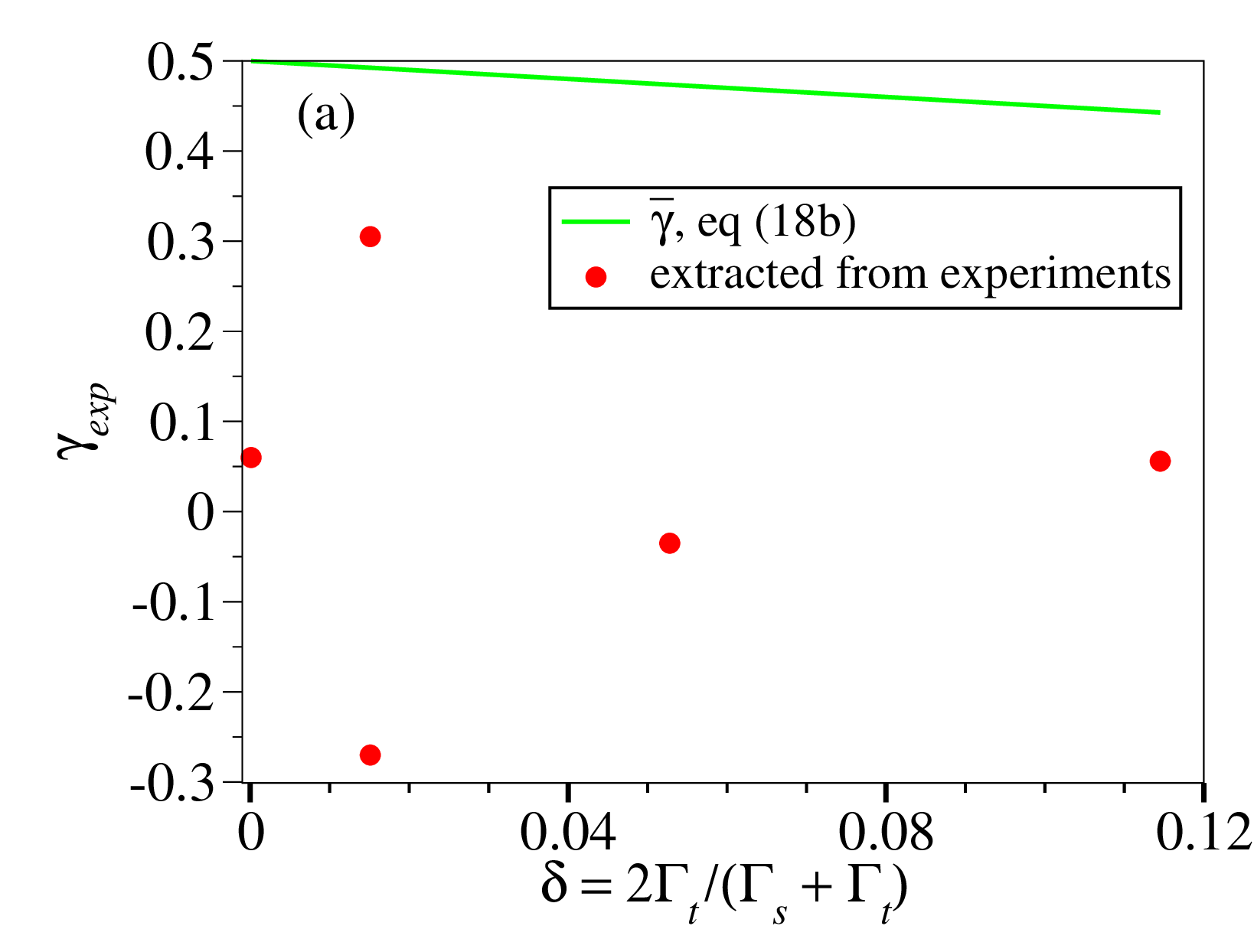}
    \includegraphics[width=0.30\textwidth,height=0.20\textwidth,angle=0]{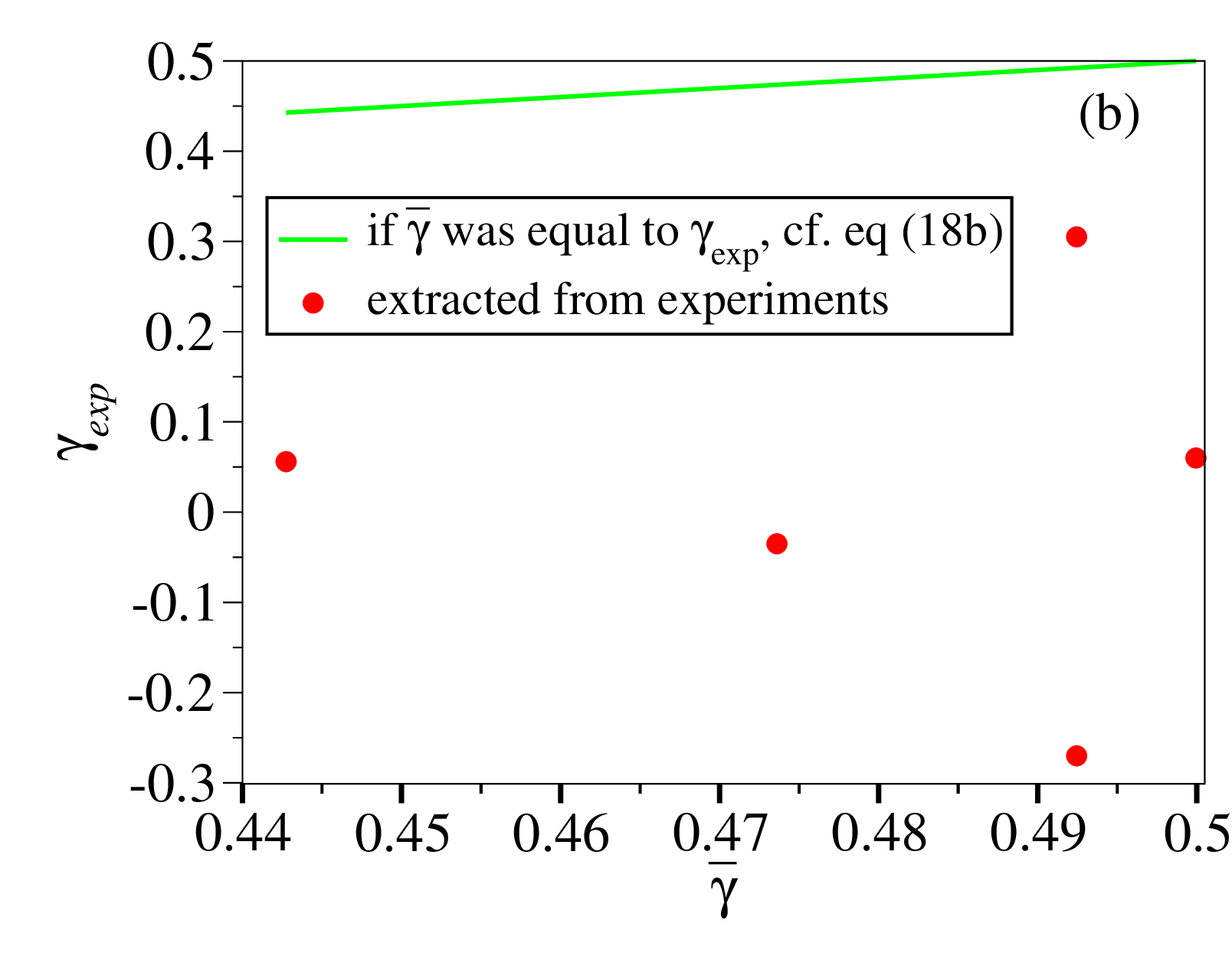}
  }
  \caption{(a) The bias-driven MO shift $\gamma_{real} \equiv \gamma_{exp}$
  deduced from transport measurements plotted against the MO coupling asymmetry parameter
  $\delta$ defined by \gl~(\ref{eq-gamma-CS}) reveals that the latter parameter has no impact on current rectification. (b) If \gl~(\ref{eq-CS}) was correct $\overline{\gamma}$
  and $\gamma_{real}$ would be equal but, as visible, they are not. Numerical values underlying this figure are collected in Table~\ref{table:gamma-delta}.}
  \label{fig:gamma-CS}
\end{figure*}

Besides the examples presented earlier,\cite{Baldea:2013d,Baldea:2014f} let us demonstrate that
the transport data for the presently considered C8T and OPT3 junctions
also invalidate \gl~(\ref{eq-CS}). Indeed, inserting the
values of $\Gamma_{s}$ and $\Gamma_{t}$ from \secname\ref{sec:c8t-opt3} into \gl~(\ref{eq-gamma-CS})
we get via $\overline{\gamma}_{C8T} = 0.47 $ for C8T and $\overline{\gamma}_{OPT3} = 0.44 $ for OPT3.
With $\varepsilon_{0}(V)$ of \gl~(\ref{eq-CS}) this translates
into $\left . \mbox{RR}\right\vert_{V=1.5\,\mbox{\scriptsize V}} = 652 (\gg 1)$ for C8T
and  $\left . \mbox{RR}\right\vert_{V=1.2\,\mbox{\scriptsize V}} = 282 (\gg 1)$ for OPT3.
These values are not only quantitatively but also qualitatively different from the experimental
values: $\left . \mbox{RR}^{exp}\right\vert_{V=1.5\,\mbox{\scriptsize V}} \simeq 0.7 < 1$\cite{Baldea:2019h}
and $\left . \mbox{RR}^{exp}\right\vert_{V=1.2\,\mbox{\scriptsize V}} \simeq 2.5 > 1$.\cite{Baldea:2019d}
For a more complete overview on the unsuitability of \gl~(\ref{eq-CS}), all present and earlier values
mentioned above are compiled in Table~\ref{table:gamma-delta} and depicted graphically in
\figname\ref{fig:gamma-CS}.

Put conversely, let us now assume that \gl~(\ref{eq-gamma-CS}) was correct
(i.e., $\overline{\gamma} = \gamma_{real}$) 
and applied to the OPT3 junctions considered above. (Remember, this cannot be done
for C8T junctions wherein $\gamma_{real} \equiv \gamma_{exp} = -0.03$ \cite{Baldea:2019h}
would imply $\Gamma_t > \Gamma_s$ and hence, completely unrealistically, a charge transfer rate
$\Gamma_s$ between the substrate and HOMO located in its close vicinity
smaller than the HOMO-tip charge transfer rate $\Gamma_t$
across the long alkyl backbone.) 
With the value $\gamma_{real} \equiv \gamma_{exp} = 0.055$ extracted from
OPT3 $I$-$V$-data,\cite{Baldea:2019d} we get
$\Gamma_{s}^{OPT3} = \Gamma^{OPT3} \sqrt{\left(1 + 2 \overline{\gamma}\right) / \left(1 - 2 \overline{\gamma}\right)} = 5.05$\,meV.
Being substantially smaller than $\Gamma_{s}^{OPD3} = 18.34$\,meV (cf.~\secname\ref{sec:c8t-opt3}),
this value $\Gamma_{s}^{OPT3} = 5.05$\,meV is unphysical; located at center of the symmetrical OPD3 molecule,
OPD3 HOMO's center-of-charge is more distant from the substrate than OPT3 HOMO's center-of-charge
displaced from the molecular center towards the thiol end.
\begin{table}
  \centering
  \caption{Values of $\gamma_{real} \to \gamma_{exp}$, $\delta$, and $\overline{\gamma}$ for several molecular junctions investigated experimentally.}
  \label{table:gamma-delta}
  \begin{threeparttable}
\begin{tabular}{rccc}
      \hline
        $\gamma_{real}$ & $\delta$ & $\overline{\gamma}$  & System \\
    \hline
    0.056 \tnote{$^a$}    &    0.115 \tnote{$^f$}   &   0.443 \tnote{$^{g}$}  &   CP-AFM, OPT3 \cite{Baldea:2019h} \\
   -0.035 \tnote{$^b$}    &    0.053 \tnote{$^f$}   &   0.474 \tnote{$^{g}$} &   CP-AFM, C8T  \cite{Baldea:2019d} \\
    0.060 \tnote{$^{c,d}$} &    1.1e-4 \tnote{$^{c}$} &   0.500 \tnote{$^{g}$} &   EC-STM (variable bias mode), azurin \cite{Artes:12b}  \\
   -0.305 \tnote{$^{e,d}$} &    0.015  \tnote{$^{e}$} &   0.492 \tnote{$^{g}$} &   EC-STM (variable bias mode), viologen \cite{Wandlowski:08} \\ 
   -0.270 \tnote{$^{e,d}$} &   0.015  \tnote{$^{e}$} &   0.492 \tnote{$^{g}$} &   EC-STM (constant bias mode), viologen \cite{Wandlowski:08} \\ 
\hline
  \end{tabular}
    \begin{tablenotes}\footnotesize
    \item[$^a$]  Ref.~\citenum{Baldea:2019h}
    \item[$^b$]  Ref.~\citenum{Baldea:2019d}
    \item[$^c$]  Ref.~\citenum{Baldea:2013d}
    \item[$^d$]  Notice that $\gamma$ of refs.~\citenum{Baldea:2013d} and \citenum{Baldea:2014f} corresponds to $\gamma_{real} = 1/2 - \gamma $ 
    \item[$^e$]  Ref.~\citenum{Baldea:2014f}
    \item[$^f$]  This work
    \item[$^g$]  This work via \gl~(\ref{eq-gamma-CS})
    \end{tablenotes}
    \end{threeparttable}
\end{table}

To avoid misunderstandings, one should finally note that in this work emphasis was on the fact that
the asymmetric coupling of the dominant level to electrodes ($\Gamma_s \neq \Gamma_t$)
does not give rise to current rectification in most real molecular junctions.
This does by no means imply that current rectification cannot be quantitatively
described within the single level model. 
Along with the imbalance between the electrodes' (electro)chemical potential $\mu_{x}$ (\gl~(\ref{eq-mu}))
an applied bias $V$ may in general yield a (bias-driven) shift of the
energy level
\begin{equation}
\label{eq-gamma-genuine}
\varepsilon_{0} \equiv \left . \varepsilon_{0}\right\vert_{V=0} \xlongrightarrow{V\neq 0} \varepsilon_{0}(V) =
\varepsilon_{0} + \gamma_{real}\, e V
\end{equation}
This was quantitatively shown in experimental data analysis,\cite{Baldea:2018a,Baldea:2019d,Baldea:2019h}
with the important observation that the above $\gamma_{real}$ turned out to be a parameter independent
of $\Gamma_s$ and $\Gamma_t$ which is neither equal to $\gamma$ entering \gl~(\ref{eq-renormalized-e0-gamma})
nor to $\overline{\gamma}$ of \gl~(\ref{eq-gamma-CS}).
It is the opposite energy shift direction caused by positive and
negative biases that gives rise to rectification, which can be accounted theoretically
by means of the single model (cf.~\gl~(\ref{eq-H}))
\begin{equation*}
\varepsilon_{0}(\ldots) \to \varepsilon_{0}(V) 
\end{equation*}
wherein the $V$-dependence is expressed by \gl~(\ref{eq-gamma-genuine}).
In general, the dependence on $V$ of
$\varepsilon_{0}(V)$ expressed by \gl~(\ref{eq-gamma-genuine})
results from the interplay between intramolecular Stark effects \cite{Baldea:2018c} and 
off-center spatial location of MO's center of charge.\cite{Metzger:01b,Datta:03,ZhangKuznetsov:08}
The latter (expression of the ``lever'' \cite{Metzger:01b} or
``potentiometer'' \cite{Baldea:2015g,Baldea:2018c} rule) results from convoluting the
MO's spatial distribution with the local electric potential whose
determination requires to simultaneously (self-con\-sis\-ten\-tly) solve the quantum mechanical (Schr\"o\-din\-ger)
and electrostatic (Poisson) equations.
\ib{Such microscopic calculations turned out to be successful in quantitatively
  reproducing RR of OPT3 junctions even subject to a mechanical deformation.\cite{Baldea:2021x}} 
State-of-the-art ab initio calculations \cite{Baldea:2015g,Baldea:2018c} showed that in alkanethiols
the intramolecular Stark effect yields a strictly linear $V$-dependence well beyond the bias range
sampled in experiments.\cite{Baldea:2018c,Baldea:2019h}

To be on the safe side, 
we wrote above that $\Gamma_s \neq \Gamma_t$ cannot yield RR-values significantly different from unity 
``in most real molecular junctions''. 
The results deduced in this work showed that this is indeed the case irrespective of
whether the charge transport is off-resonant (cf.~\gl~(\ref{eq-off-resonance})) or on-resonant.
In situations escaping the model of \gl~(\ref{eq-H}) --- unlikely in real molecules but still possible in artificial
nanostructures where electronic properties can be continuously tuned --- close to resonance,
level's occupancy may be significantly different from zero and $V$-dependent.
\cite{Baldea:2014f,Baldea:2015a,Baldea:2015c,Baldea:2015g} If furthermore 
on-site electron-electron interactions \cite{Stafford:96,Buttiker:03}
or electron-electron interactions at contacts \cite{Baldea:2012b}
are strong,  $\Gamma_s \neq \Gamma_t$ may lead to a certain $I$-$V$ asymmetry,
although spectacular RR-values can hardly be expected on this basis. Problems arise
in those cases
because, close to resonance, electron couplings both to slow vibrational degrees of freedom (reorganization effects)
fast phonons (deserving quantum-mechanical treatment) need be considered.
\section{Conclusion}
\label{sec:conclusion}
\ib{It was not our aim here to explain current rectification in real molecular junctions using schematic (tight binding, Hubbard, etc)
  models extended, e.g., to also include interactions due to charge accumulation at interfaces
  considered in \secname\ref{sec:occupancies}. Realistic microscopic calculations showed that RR can quantitatively described in real molecular junctions
  even subject to mechanical stretching \cite{Baldea:2021x} or is situations where, counterintuitively,
  the dominant MO does not track the substrate in its close proximity but rather the much more distant tip electrode.\cite{Baldea:2015g,Baldea:2018c}}

Rather, in this work, we have presented analytic results deduced theoretically by exactly solving
the nonequilibrium problem for a general quantum mechanical Hamiltonian describing the charge transport
dominated by a single energy level.
Technically speaking, the present study goes beyond the existing approaches to charge transport and related current rectification
within a single dominant channel because we worked out the general equation \gl~(\ref{eq-I})
for the current by employing exact expressions for the embedding self-energies having: (i) nonvanishing
real parts (cf.~\gl~(\ref{eq-Delta_x-ex})) and (ii) imaginary parts that do depend on energy (cf.~\gl~(\ref{eq-Gamma_x-ex})). 

The formulas deduced in this way enabled us to obtain numerical estimates based on parameter values both
extracted from transport measurements on benchmark junctions or
even chosen to simulate rectification enhancement.
On this basis, we can definitely rule out that unequal MO couplings to electrodes ($\Gamma_s \neq \Gamma_t$)
make a significant contribution to current rectification in experiments with molecular junctions fabricated so far.
This conclusion clearly contradicts some opposite claims in previous literature while confirming other assertions
based on intuitive considerations.\cite{Datta:97a,Hush:00}

\emph{Hypothetically},
$\Gamma_s \neq \Gamma_t$ could yield (albeit not large but still presumably) observable $I$-$V$-asymmetry
in case of electrodes possessing extremely narrow conduction
bands (low $W_{s,t}$ imply large $\gamma$, cf.~\gl~(\ref{eq-renormalized-e0-gamma})).
Artificial nanostructure may be better suited for this purpose because their properties can be tuned easier than
those of real molecules.
Still, in those cases electron correlations will certainly be very strong
and invalidate (Laudauer's) uncorrelated transport description underlying
the vast majority of theoretical studies including the present one.
\section*{Acknowledgements}
The author gratefully acknowledges financial support from the German Research
Foundation (DFG grant BA 1799/3-2) in the initial stage of this work and computational support
by the state of Baden-W\"urttemberg through bwHPC and the German Research Foundation through
grant no INST 40/575-1 FUGG (JUSTUS 2 cluster).
%

%
\end{document}